\documentstyle[prb,aps,psfig]{revtex}
\begin{document}
\draft
\title{The random-field specific heat critical\\
behavior at high magnetic concentration: $\rm Fe_{0.93}Zn_{0.07}F_2$}

\author{Z. Slani\v{c} and D. P. Belanger}
\address{Department of Physics, University of California\\
Santa Cruz, CA 95064, USA}
\date{\today}
\maketitle
\begin{abstract}
The specific heat critical behavior is measured and analyzed
for a single crystal of the random-field Ising system
$\rm Fe_{0.93}Zn_{0.07}F_2$ using pulsed heat and optical
birefringence techniques.  This high magnetic concentration
sample does not exhibit the severe scattering hysteresis at low
temperature seen in lower concentration samples and its
behavior is therefore that of an equilibrium
random-field Ising model system.  The equivalence of the behavior
observed with pulsed heat techniques and optical birefringence
is established.  The critical peak appears to be a symmetric,
logarithmic divergence, in disagreement with random-field model
computer simulations.  The random-field specific heat scaling
function is determined.
\end{abstract}


\section{Introduction}

It has been known experimentally and theoretically
for several years that a phase transition
occurs \cite{by92} for the three-dimensional ($d=3$)
random field Ising model (RFIM).  The first evidence \cite{bkjc83}
for the phase transition came from the 
measurement of ($d(\Delta n)/dT$), where $\Delta n$ is the optical 
linear birefringence
and $T$ is the temperature.
It has long been held \cite{fg84,bkj84}
that $d(\Delta n)/dT$ is proportional
to the magnetic component of the specific heat in anisotropic
antiferromagnets to a high degree of accuracy.  This has been shown
explicitly \cite{bnkjlb83} for the pure three dimensional ($d=3$)
antiferromagnets $\rm MnF_2$ and $\rm FeF_2$ and
for the magnetically dilute system \cite{db89} $\rm Fe_{0.46}Zn_{0.54}F_2$
using optical and pulsed heat techniques.
The specific heat critical behavior can be obtained in
principle from either technique.  In applicable cases, however,
the birefringence
technique is preferable for two reasons. First, the non-magnetic
contribution to $d(\Delta n)/dT$ is insignificant, whereas
the non-magnetic phonon background for the pulsed heat technique is
often large and difficult to eliminate for the purposes of analyzing
the magnetic critical behavior.  Second, by aligning the probing
laser beam to be perpendicular to the inevitable concentration
gradients in dilute crystals, one can greatly minimize the effects of
the concentration variations that often mask the true critical
behavior \cite{kfjb88}.  Thus, using the optical technique allows
data to be taken much closer to the transition.
For measurements in an applied field $H>0$, the birefringence method
works well for measuring magnetic specific heat critical behavior,
as should techniques that are sensitive to the uniform
magnetization, e.g.\ magnetometry or optical Faraday rotation.
The universal critical exponent $\alpha$ and amplitude ratio
$A^+/A^-$ for the specific heat for $H>0$ can be measured with any
one of these techniques as well as with the pulsed heat technique.
The only difference is in the field dependence of the amplitudes.

The scaling behavior of the amplitudes
for dilute antiferromagnets has been worked out by Kleemann,
et al.\ \cite{kkj86} by considering leading singularities in
derivatives of the free energy for $H > 0$.
The magnetic specific heat for small fields near the phase transition
has the dependence
\begin{equation}
C^m = h_{RF}^{(2/\phi)(\alpha - \alpha ^*)}g(|t_h|h_{RF}^{-2/\phi}) \sim h_{RF}^{(2/\phi)(\alpha - \alpha ^*)}|t|^{-\alpha}
\end{equation}
where $t_h = (T-T_N +bH^2)/T_N$, $T_N$ is the zero field transition
temperature, $T_c(H)$ is the transition temperature in the field,
b is a mean-field temperature shift coefficient,
$t=(T-T_c(H))/T_c(H)$ is the reduced temperature,
$\alpha ^*$ is the $H=0$ random-exchange specific heat
critical exponent, $\alpha$ is the $H>0$
random-field exponent, $h_{RF}$ is the random-field
strength which is proportional \cite{c84} to the applied field $H$,
and $\phi$ is the random-exchange to random-field crossover exponent.
Note that since our maximum applied field $H=7$~T is much smaller
than the spin-flop field\cite{jkmsd83} $H_{SF} = 41.9$~T for $FeF_2$, the
demagnetization field corrections are negligible.
Similarly, the uniform magnetization measurements yield the
scaling behavior
\begin{equation}
\left( \frac {\partial M}{\partial T}\right) _H \sim h_{RF}^{(2/\phi)(1+\alpha - \alpha ^*- \phi /2)}|t|^{-\alpha}
\end{equation}
and
\begin{equation}
\left(\frac {\partial M}{\partial H}\right) _T \sim h_{RF}^{(2/\phi)(2+\alpha - \alpha ^* -\phi)}|t|^{-\alpha} \quad .
\end{equation}
The Faraday rotation yields the same scaling behavior as the uniform
magnetization.  $d(\Delta n)/dT$ is expected to
yield the scaling behavior of the magnetic specific heat, $C^m$.
The proportionality between $C^m$ and $d(\Delta n)/dT$
is valid in the critical region for very anisotropic antiferromagnets.
This is true to a high degree of accuracy \cite{rkj88}
for $\rm Mn_xZn_{1-x}F_2$ for
all concentrations that show critical behavior.  The order of magnitude larger
anisotropy of $\rm Fe_xZn_{1-x}F_2$ ensures even greater accuracy.

The primary motivation for returning to the study of the
specific heat critical behavior of dilute anisotropic antiferromagnets
is to characterize the universal parameters of the
random-field Ising model (RFIM) in a system which shows
equilibrium behavior below $T_c(H)$.
With a small applied field along the
easy axis, dilute anisotropic antiferromagnets are proposed
as ideal examples of the RFIM, the same critical behavior
universality class as a ferromagnet with an applied random field \cite{c84}.
Since most computer simulation studies of the critical behavior
have been made using the latter model,
the relevance of the comparison between the experiments and simulations
rests on the universality of the behavior of the two systems.
However, the correspondence has been far from obvious in most cases
since the experiments have been done using crystals with $x<0.80$.
Low temperature hysteresis in the scattering line shapes
and anomalous behavior of the
Bragg scattering intensity versus $T$ in these crystals have prevented
a comprehensive characterization of the random-field critical behavior,
particularly for $T<T_c(H)$.  In particular, no long-range order is
observed after cooling in a field for $x<0.80$.
Although the applied fields are small enough to ensure proper comparison
with theory valid in the small random-field limit,
the dramatic role of the numerous vacancies at these concentrations had not
been fully appreciated until recently.  When the number of
vacancies is too large, domains can form with little energy cost
since the walls pass predominantly through the vacancies \cite{nu92,hb92}.
Studies \cite{bkjn87,bwshnlrl95} on the bulk crystal
$\rm Fe_{0.46}Zn_{0.54}F_2$ and
the epitaxial film $\rm Fe_{0.52}Zn_{0.48}F_2$ have demonstrated this.
In contrast, the crystal used in the present
experiments, with $x=0.93$, shows no anomaly in the Bragg scattering
intensity and no low temperature hysteresis in the
scattering line shapes \cite{sbf97}.
Unlike the lower concentration samples, long-range order is observed
upon cooling in an applied field.
This is attributed to the low vacancy density and the corresponding
energy cost of domain formation.  Only at high magnetic concentration
might the Imry-Ma domain arguments \cite{im75} be applicable.
Hence, the $\rm Fe_{0.93}Zn_{0.07}F_2$
crystal affords us the opportunity to characterize the random-field
critical behavior in a system that exhibits equilibrium behavior
below the transition.
We characterize the specific heat critical behavior
of this sample and make comparisons to earlier studies \cite{db89} at lower
concentration.
The fact that the crossover from REIM to RFIM behavior occurs
within the critical region, i.e.\ for $|t|<<0.1$ allows us to also determine
the crossover function $g(|t_h|h^{-2/\phi})$ in Eq.\ 1.

A secondary reason to examine this sample, and in particular
to do pulsed specific heat as well as birefringence
measurements, is to verify the proportionality of
the data obtained from these two techniques since this has
been questioned \cite{bfhhrt95} in recent series of letters
and review articles
proposing the so-called ``trompe l'oeil'' phenomenological model.
The model attempts to explain the anomalous behavior of the Bragg
scattering $\rm Fe_{0.5}Zn_{0.5}F_2$ and $\rm Mn_{0.75}Zn_{0.25}F_2$
as a violation of RFIM scaling laws.  The model requires
that $C^m$ and $d(\Delta n)/dT$ (or $dM/dT$)
exhibit different critical behavior in dilute antiferromagnets
once a field is applied and, in fact, to exhibit peaks at
different temperatures.  We can use the measurements on 
$\rm Fe_{0.93}Zn_{0.07}F_2$ along with previous ones on
$\rm Fe_{0.46}Zn_{0.54}F_2$ to test this proposition.
We will examine whether the data are consistent with conventional
RFIM scaling theory properties and the formation of domains at low
temperature or, rather, with the alternative
proposed ``trompe l'oeil'' phenomenology.

\section{Experimental Details}

Pure $\rm FeF_2$ is an excellent Ising system owing
to its large anisotropy and dominating second-nearest
neighbor interaction \cite{by92}.  Mixed crystals of $\rm FeF_2$ and
$\rm ZnF_2$ can be grown with high crystalline quality, very
small concentration gradients and no indication of chemical
clustering.  The Ising character is preserved
upon dilution.  The $\rm Fe_{0.93}Zn_{0.07}F_2$ crystal
used in the pulsed heat measurements exhibits a rounding of the
transition temperature for $|t|< 2 \times 10^{-3}$
due to the concentration variation
determined from room temperature $\Delta n$ measurements \cite{kfjb88}.
The birefringence data are rounded only
for $|t|<2 \times 10^{-4}$ since the
probing laser is oriented in a direction
perpendicular to the gradient.

The boule from which the $\rm Fe_{0.93}Zn_{0.07}F_2$ crystal was cut was
grown \cite{nk} at the University of California, Santa Barbara.
The concentration gradient profile was measured using room
temperature birefringence and the $1.345 \; \rm g$ sample was
cut from the portion of greatest homogeneity.  The magnetic concentration
was determined from the zero field transition temperature which is
linear \cite{by92} for $0.4<x<1.0$.

For the $\Delta n$ measurements,
two parallel faces were polished parallel to the c-axis.
The laser beam ($\lambda = 632.8 \; \rm nm$) traverses a sample
thickness of $3.27 \; \rm mm$.  The sample was oriented with its
c-axis parallel to $H$. A $0.5 \; \rm mm$ pin-hole in front
of the sample minimizes the effects of concentration gradients
and vibrations.  The S\'{e}narmont technique is used
to measure $\Delta n$ with a resolution of
$2 \times 10^{-9}$.  
A commercially calibrated carbon resistor thermometer was 
chosen for temperature
measurement and control because of its low field dependence and high
sensitivity. It was possible to achieve temperature stability better than
$50 \; \rm \mu K$. The same thermometer was later used in the specific heat
measurement and was then also calibrated in field vs.\ another thermometer,
whose field dependence is known (see below).

Three different thermal procedures were employed.  For
zero-field-cooling (ZFC), the sample is first cooled below $T_c(H)$
with $H=0$, the field is subsequently raised
and the sample is slowly heated through $T_c(H)$
in small temperature steps.
Specific heat measurements show that the sample
temperature usually stabilizes after about $120 \; \rm s$ after 
the application of the heat pulse.
The temperature in the birefringence measurements 
is typically stabilized at each temperature step
for approximately $400 \; \rm s$ before a $\Delta n$ measurement is taken,
thereby giving the sample ample time to come to equilibrium.
For field-cooling (FC), the
field is raised well above $T_c(H)$ and the sample is
cooled through $T_c(H)$ as data are taken in the manner
described above.  For field heating (FH),
the sample is first FC and then data are taken while the temperature
is increased.  The temperature steps in all
cases were approximately $0.5 \; \rm K$ far away from the
transition decreasing to $0.005 \; \rm K$ close to $T_c(H)$.
Reasonable variations in the rates of heating and cooling and stabilization
times had no observable effect on the data.

For the adiabatic heat pulse technique, the sample chamber is mounted onto
the cold finger via a narrow copper neck. A thermal shield surrounds
the sample chamber and the neck. It
is fixed to the cold finger and is therefore at the same
temperature as the cold finger, i.e.\ at the temperature of the nitrogen bath.
A heater wire, controlled by the 
temperature controller, is wrapped tightly around and varnished onto the neck
below the mounting point for the thermal shield. Two carbon resistor
thermometers are mounted into the neck below the heater but above the
sample chamber. It is assumed that the neck below the heater, the two
thermometers in the neck and the sample chamber are at the same
temperature, controlled by the heater on the neck.

The sample is mounted on a thin sapphire plate using GE7031
varnish along with a small Stablohm 800 wire heater.
The heater is connected to a four-wire
constant-power supply.  An unshielded carbon thermometer
is attached with varnish to the sample and is connected
using a four-wire technique to a current ratio transformer bridge.
An unshielded thermometer was chosen to minimize
the thermometer's specific heat contribution.
The sample is suspended inside the sample chamber by the $0.0254 \; \rm mm$
Cu wires used for the thermometer and heater.
Care is taken to ensure that the c-axis of the sample is parallel to $H$.
The wires provide a small heat leak from the sample controlled
by the temperature difference, $\delta T$,
between the sample chamber and the sample.  The sample thermometer
produces a small amount of heat which is compensated by the
controlled heat leak so that, in the absence of the heat pulse, the
sample temperature is constant.  One thermometer
located in the neck above the sample chamber 
is connected to the same bridge as the sample thermometer and is used
to control $\delta T$.  A second thermometer in the neck is
used with a second bridge to determine the absolute temperature of the neck 
and
is commercially calibrated for $H=0$. This thermometer is the same as the
one used in the optical birefringence measurement.
A preliminary experiment is performed to calibrate $\delta T$ versus $T$
for sample temperature stability.  The calibration is incorporated
into the computer control algorithms so that the balance is automatically
preserved over the entire temperature range during the specific heat
experiment.

During a ZFC experimental run, the sample temperature is
first lowered well below $T_c(H)$.
The field is then raised. An equilibrium
temperature difference between the sample and the sample chamber is found
with a temperature drift of the sample of $50 \; \rm \mu K$ per minute or less.
The absolute temperature of the sample chamber is measured and a heat
pulse is applied.  The total heat pulse
is controlled by the duration of the $907 \; \rm \mu W$ pulse.
A typical $18 \; \rm mJ$ pulse results in a temperature change of approximately
$0.025 \; \rm K$, corresponding to a change of $3\times 10 ^{-4}$ in
reduced temperature, $t$, close to $T_c(H)$.
As the temperature increases, $\delta T$
is kept constant, so the temperature of the sample chamber rises
as the sample heats.
After the application of the heat pulse, the sample is given $300 \; \rm s$
to equilibrate and the temperature of the sample chamber is
recorded.  $\delta T$ is then set to the new correct value
corresponding to the new temperature.  The next heat pulse is then applied.

A FC technique completely equivalent to the one used in birefringence is
difficult since it is difficult 
to extract a known amount of heat from the sample. Instead, for the FC
technique, the sample temperature is set to a value well above $T_c(H)$.
The sample chamber temperature is then lowered by turning
off the sample chamber heater, thereby causing the sample
and chamber to slowly cool.  After
approximately $15$ to  $20 \; \rm min$, the heater is turned on and
the sample chamber temperature is increased.
The sample chamber temperature is then measured
and $\delta T$ is adjusted alternately until the sample is equilibrated.
A heat pulse is applied to the sample and the
change of temperature is measured as described for ZFC.
The sample temperature is then lowered again.
Using reasonably different-sized cooling steps did not significantly
affect the data.

After collecting all the data, the sample is removed from the sample
chamber and the thermometer is fixed to the sapphire plate.  The specific
heat of the thermometer, the sapphire plate, the varnish and the wires is
then measured at $H=0$. This background specific heat is subtracted from all
of the specific heat data before further analysis.

To calibrate the thermometer field dependence in the temperature region of the
transitions, the temperature of the sample is stabilized near 
and below where the transition takes place in 7 Tesla. The field is then
raised to 3 Tesla and the new resistance of the sample thermometer is
measured. The field is then decreased at the same rate as it was increased
before. The sample thermometer resistance is measured again. If one assumes
that the sample was heated equally during the field decrease and
increase, then the resistance measured at 3 Tesla
corresponds to the mean value of the initial and final resistance in zero
field. The thermometer is therefore calibrated at that temperature for 3
Tesla. The procedure is then repeated for 5 and 7 Tesla. A cubic spline
interpolation is used to find the temperature (or resistance) corrections
for all the other fields.  The whole
process is repeated at a temperature near and above the zero field
transition. The field corrections for all the temperatures between the two 
calibrated temperatures is arrived at by the use of linear interpolation 
between the calibrated temperature points. 
Since the field corrections are small, this procedure yields sufficient
accuracy.

At the end of the experiment, the sample thermometer was mounted onto the 
sample chamber by GE7031 varnish. The uncalibrated thermometer in the  neck
and the sample thermometer were then calibrated against the commercially 
calibrated thermometer in the neck. Afterwards, all the thermometers were
also calibrated in field. 
It was possible to determine field corrections for the two thermometers in
the neck, since the behavior of the sample thermometer in the field was 
known.

It should be noted that the unshielded thermometer shows significant
variations in resistance after the temperature is cycled to room
temperature and cooled again. If one uses the calibration obtained in 
the process described in the previous chapter, the birefringence and the
specific heat data do not coincide, even in zero field.  The two
shielded thermometers in the neck, however, behaved very consistently and
within the specified tolerances.  The sample thermometer was
recalibrated after every cycling using the sample's zero 
field transition temperature,
for which there is no ambiguity. The measured resistance was multiplied by
the required factor so that the birefringence and the specific heat data
zero field transition temperatures agreed. The changes in the resistance
vs.\ temperature curves for the sample thermometers probably come from
induced strain in the thermometer when it is cycled to the room
temperature.

\section{Results and Discussion}

The magnetic specific heat behaviors measured using
the birefringence technique with
$H=0$, $5$ and $7 \; \rm T$ are shown in Fig.\ 1.  In the main figure we
show $H=0$ data and, for $H>0$, ZFC data.
In the inset we show both the ZFC and FC
data for $H=7 \; \rm T$ demonstrating the hysteresis extremely
close to $T_c(H)$.
The curves in this figure are not fits to the data but simply
represent the smoothed behavior of the data.
Just as observed at lower concentrations,
FC yields a more rounded behavior close to $T_c$.

Figure 2 shows the magnetic specific heat $C^m$, obtained from
the pulsed heat data $C_p$ by subtracting the nonmagnetic
background.  We are able to do the background subtraction to a good
degree of accuracy by first determining the background at $H=0$.
This is done by comparing the $d(\Delta n)/dT$ and
$C_p$ data, assuming the $d(\Delta n)/dT$ nonmagnetic
background is negligible and that $d(\Delta n)/dT$ and $C^m$
are proportional at $H=0$.  The excess signal found in the
$C_p$ data is then taken to be the nonmagnetic contribution
to the specific heat, an otherwise difficult quantity to
obtain.  Within the accuracy of the measurements, the
proportionality between $d(\Delta n)/dT$
and $C^m$, $A=9.17 \times 10^{-6}$, is the same as that obtained
for pure $\rm FeF_2$ and $\rm Fe_{0.46}Zn_{0.54}F_2$, i.e.\ the proportionality
is independent of the concentration.  Once the nonmagnetic contribution
to the specific heat is determined, it is subtracted from all of
the specific heat data sets.  The ZFC data in the main figure are
more rounded than the corresponding $d(\Delta n)/dT$ data shown
in Fig.\ 1, a consequence of the
greater variation in the concentration gradient across the sample
since the whole crystal is used in the $C_p$ measurement.
The solid curves in the main figure are obtained from
the ZFC $d(\Delta n)/dT$ data by numerically rounding
a cubic spline fit of the data to match the
concentration variation in the specific heat sample and
then drawing smooth curves through the rounded spline fit.
The rounded curves are then
transferred to the ZFC data in Fig.\ 2.  No other adjustments are
made.  The data are clearly well represented by the curves, demonstrating
that the difference in the appearance of the ZFC $d(\Delta n)/dT$
and $C^m$ data is attributable entirely to the concentration gradient
in the sample.  The FC behavior at $H=7 \; \rm T$ is shown in the inset of
Fig.\ 2.  The curves are taken from the $d(\Delta n)/dT$
data in the same manner as described above.  The FC curve describes
the data well, again showing that the $d(\Delta n)/dT$ and
$C^m$ data are clearly proportional, once one properly accounts for
the gradient effects.  The dotted curve is the
one corresponding to the ZFC data and is the same as the solid curve
in the main figure.  The ZFC data are not shown in the inset simply
to preserve clarity, but the agreement with the ZFC curve and data is
evident from the main figure.  The fact that there is only a
small difference between the ZFC and
FC $C^m$ behavior, relative to the $d(\Delta n)/dT$ data, is
simply a consequence of the concentration variation sampled with the
$C^m$ technique.
Experiments such as the one \cite{bfhhrt95} performed on
$\rm Fe_{0.5}Zn_{0.5}F_2$
which do not exhibit hysteresis most likely suffer from concentration
gradients and should not be taken as evidence that hysteresis is
absent.  Indeed, the $\rm Fe_{0.46}Zn_{0.54}F_2$ crystal, which is of
superb homogeneity with $\delta x < 2 \times 10^{-4}$, clearly shows hysteresis
in the specific heat \cite{db89}.

We do not show the FH $d(\Delta n)/dT$ data since
they behave nearly indistinguishably from the ZFC data.
This implies that once the sample is FC to low temperatures, the
resulting state must be extremely close to that of the ZFC one.

Since we have clearly established that $d(\Delta n)/dT$ and $C^m$
experiments yield the same critical behavior for all $H$, we will
concentrate on
the former, which are less affected by gradients and have less ambiguity
associated with the nonmagnetic background.  Figure 3 shows
$d(\Delta n)/dT$ versus the logarithm of $t$
for $H=0$, $2$, $5$ and $7 \; \rm T$.  
For $H=0$, the data are consistent with an asymmetric cusp,
the expected random-exchange critical behavior.  However, directly
fitting the data did not give as reliable an exponent as we were
able to infer from the scaling behavior to be discussed below. 
The ZFC critical behavior at $H=7 \; \rm T$ is quite different.
We do not observe
significant rounding for the fields and reduced temperatures accessed.
For $|t|<2.5 \times 10^{-3}$, the data for $T>T_c(H)$ and $T<T_c(H)$
lie on the same
straight line, indicating a symmetric logarithmic divergence
(i.e.\ $\alpha = 0$).  At lower fields the data
also exhibit the symmetric logarithmic divergence, but over a smaller
range in $|t|$, a consequence of crossover.
The logarithmic peak has been observed previously \cite{pkb88}
using Faraday rotation in
$\rm Fe_{0.47}Zn_{0.53}F_2$ in a range of $|t|$ extending to much larger values
for comparable applied fields.  However, in this case rounding occurs
at very small $|t|$, which has been attributed to activated dynamic critical
behavior.  Such rounding is not observed in $\rm Fe_{0.97}Zn_{0.07}F_2$.  This
may be a result of using fields too small to observe dynamic rounding
for $|t|>10 ^{-4}$.  Interestingly, the FC data also show
the symmetric logarithmic behavior and the same crossover behavior
with the only difference being the rounding at very small $|t|$.
Hence, it is probable that the symmetric logarithm reflects
equilibrium behavior and FC simply superimposes rounding on top of it.
The FC rounding is consistent with the
incomplete development of long-range order as a result of
the activated dynamics and the associated logarithmic
relaxations near $T_c(H)$.
Since the $x=0.93$ sample shows no low temperature hysteresis in
the scattering line shapes, in contrast to those at lower concentrations,
and the specific heat nevertheless shows the same logarithmic divergence,
we can conclude that the nonequilibrium behavior observed in the
lower concentration samples has little effect on the specific heat
critical behavior.  This is consistent with the fact that the
specific heat involves primarily short range fluctuations, whereas the
scattering is sensitive to longer-range correlations, which are
more affected by domain formation.

The logarithmic divergence is not in agreement with computer simulation
results, though many of the other exponents obtained from scattering
experiments are to a reasonable extent in agreement.  Rieger \cite{r95}
obtains the simulation result $\alpha \approx -0.5$, for example,
compared with our $\alpha \approx 0$.  Nevertheless, the symmetric
logarithm is a feature observed in all experiments accurate enough
to probe the critical behavior from $x=0.46$ to $0.93$.

For the range of field accessible in the present experiments, the crossover
from random-exchange to random-field critical behavior in
$\rm Fe_{0.93}Zn_{0.07}F_2$ takes place well within the critical region,
i.e.\ for $|t|<<0.1$.  It is therefore
feasible to study the crossover function as in Eq.\ 1 without influences
from crossover to mean-field or any other kind of behavior.
Fig.\ 4 shows all of the ZFC birefringence data properly scaled.
The solid curves indicate the $\log _{10}|t|$ asymptotic behavior.  The scaled
data $h_{RF}^{2\alpha ^*/\phi}C^m/R$, however, follow the behavior of
the full scaling
function over a wide range of the variable $th_{RF}^{-2/\phi}$.
The $H=7 \; \rm T$ data cover the range $|t| < 0.1$, while the other
sets of data cover correspondingly smaller ranges in $|t|$.
The data sets collapse in a satisfactory way except for the lower
field ones close to $T_c(H)$.  A simple simulation of the
effect of the concentration gradient, however, clearly establishes
that near $T_c(H)$ the data should be suppressed, with the
effect being more noticeable at low fields and with a rounding roughly
following the observed effect.  In this scaling, we have assumed
$\alpha = 0$.  The value of the REIM exponent $\alpha ^*$ which
yeilds the best scaling collapse is $\alpha ^*= -0.10 \pm 0.02$, in
excellent agreement with an earlier estimate $\alpha ^*= -0.09 \pm 0.03$
taken from direct analysis \cite{bcsybkj83} of the critical behavior in
$\rm Fe_{0.6}Zn_{0.4}F_2$.  Note that the scaling of
$d(\Delta n)/dT$ as predicted for the specific heat
clearly supports the equivalence of the data from
the optical and pulsed heat techniques within a simple proportionality.

Recently, the local mean-field simulation technique has been used
by Raposo and Coutinho-Filho to investigate the concentration
dependence of domain formation in the dilute antiferromagnets\cite{rc96}.
Preliminary results indicate that domain formation does occur
only below a critical concentration, which increases with
the application of an external magnetic field.  Such studies will
help to elucidate the nature of the experiments at low and high
magnetic concentration.

The results of this study are not consistent with the interpretation
of the anomalous Bragg scattering at low magnetic concentrations
which has been proposed \cite{bfhhrt95} in a series of
papers by Birgeneau, et al.\ and called
``trompe l'oeil'' behavior.
Two propositions must necessarily hold true for the
``trompe l'oeil'' phenomenological model to adequately account
for the data.  First, the $d(\Delta n)/dT$
data must not exhibit the specific heat behavior.  In fact,
the peaks of the specific heat for $H>0$ must not coincide in
temperature with the peaks in $dM/dT$ or
$d(\Delta n)/dT$, though at $H=0$ they must.
The ``trompe l'oeil'' experimental reports do
argue that the peaks occur at different temperatures, but
the temperatures of the data were shifted arbitrarily \cite{bkm96}.
Second, the specific heat must not show the FC/ZFC hysteresis
near $T_c(H)$ that is seen in the other measurements.
Neither of these points are satisfied by measurements taken
using $\rm Fe_{0.46}Zn_{0.54}F_2$ or $\rm Fe_{0.93}Zn_{0.07}F_2$.
Hence, the high resolution, low concentration gradient experiments
simply do not support the proposed nonscaling ``trompe l'oeil'' model.

We have presented a study of the specific heat critical behavior
of the RFIM system which establishes that the pulsed specific
heat and $d(\Delta n)/dT$ yield the same behavior.
We have demonstrated that the RFIM specific heat critical behavior obeys the
predicted \cite{kkj86} scaling behavior.  We have shown that the critical
behavior of the specific heat consists of an asymptotic, symmetric, logarithmic
divergence to a very good approximation, in disagreement with
computer simulations \cite{r95} indicating a nondivergent specific heat.
A companion study \cite{sbf97} of the scattering in the same crystal
complements this study in an effort to provide a comprehensive
characterization of the static equilibrium critical behavior of the RFIM.

This work was funded by Department of Energy Grant No.\ DE-FG03-87ER45324.
We thank Ernesto Raposo for interesting discussions.

\begin{figure}[t]
\centerline{\hbox{
\psfig{figure=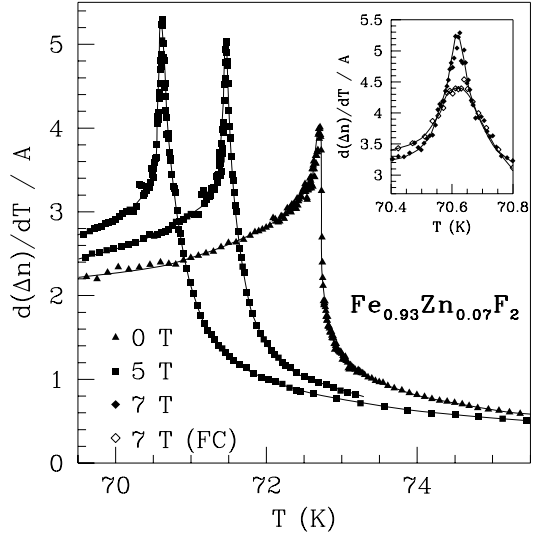,height=3.in}
}}
\caption{$ d(\Delta n)/AdT$ vs.\ $T$ for $Fe_{0.93}Zn_{0.07}F_2$.
$A=9.17 \times 10^{-6} \; \rm K^{-1}$ is the same proportionality constant
found between $C_m/R$ and $ d(\Delta n)/dT$ for pure $\rm FeF_2$.
ZFC data are shown in the main figure.
The inset shows the $H=7 \; \rm T$ FC data as well as the
ZFC data.  The curves are simply drawn smoothly through the data.
}
\end{figure}
\begin{figure}[t]
\centerline{\hbox{
\psfig{figure=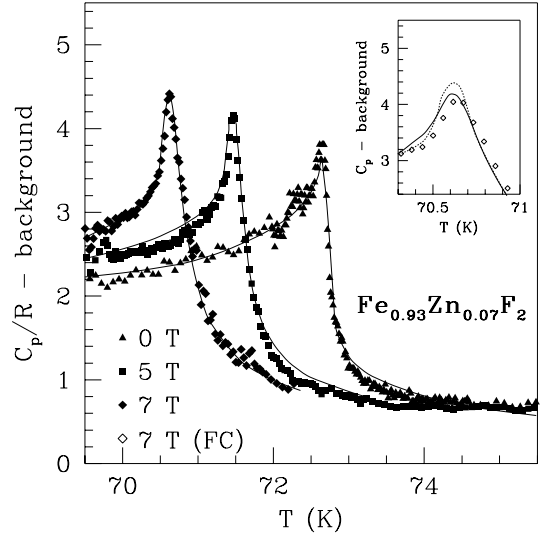,height=3.in}
}}
\caption{The magnetic specific heat $C_m=C_p/R- \rm background$ vs.\ $T$
for $\rm Fe_{0.93}Zn_{0.07}F_2$.  The phonon contribution
to the specific heat has been subtracted, as discussed in the text.
ZFC data are shown in the main figure.  The inset shows the
$H=7 \; \rm T$ FC data.  The curves
are the same as those in the previous figure except
that they are rounded by the larger
variation of the concentration gradient, as discussed in the text.
The ZFC data are not shown in the inset, for clarity,
but the dotted line is the same as the solid ZFC line in the main figure.
}
\end{figure}
\begin{figure}[t]
\centerline{\hbox{
\psfig{figure=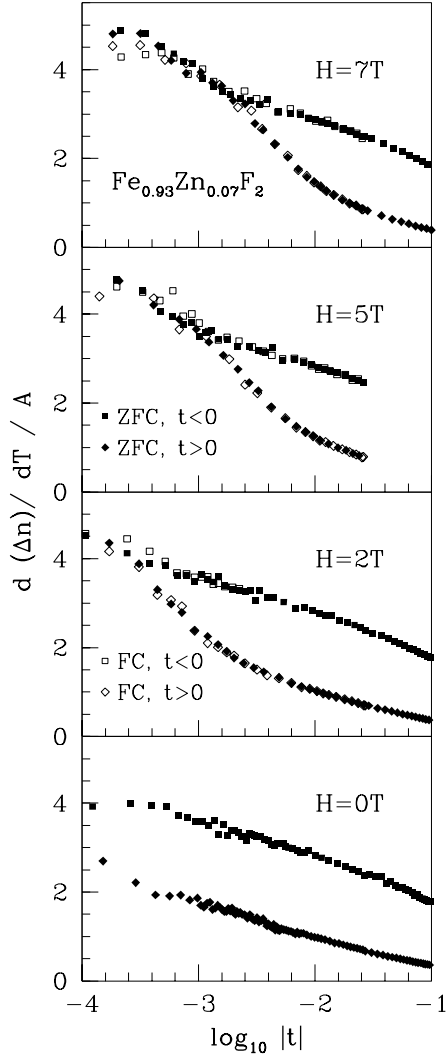,height=6.in}
}}
\caption{$d(\Delta n)/AdT$ = vs.\ $\log_{10}|t|$ for $H=0$,
$2$, $5$ and $7 \; \rm T$.  The solid symbols are for ZFC and the open
symbols are for FC.  For $H=0$, the data above and below $T_c(H)$
have distinct amplitudes.  In contrast, for $H=7 \; \rm T$, the ZFC amplitudes
are equal and the data follow an approximate straight line on
the semi-log plot for small $|t|$, indicating a symmetric
logarithmic divergence.  The FC data show significant rounding,
but the symmetric logarithmic behavior is still evident.
For the smaller fields, the crossover to the symmetric logarithmic
behavior occurs much closer to $t=0$, as expected from scaling.
}
\end{figure}
\begin{figure}[t]
\centerline{\hbox{
\psfig{figure=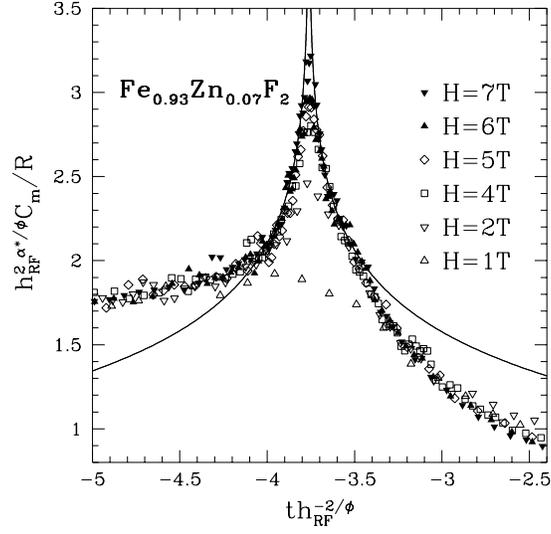,height=3.in}
}}
\caption{$ h_{RF}^{2\alpha ^*/\phi}C_m /R$, using normalized ZFC data
obtained  with the $d(\Delta n)/dT$ technique,
vs.\ $t h_{RF} ^{-2/\phi}$
for $H=1$, $2$, $4$, $5$, $6$ and $7 \; \rm T$.  The solid curves
represent $h_{RF}^{2\alpha ^* /\phi}\ln |t|$,
which the data follow in the asymptotic region.
The data collapse onto the scaling function in Eq.\ 1, except for
data close to $T_c(H)$, where rounding depresses the peaks.
The peak depression is more pronounced at lower fields, as expected,
and is consistent with the concentration variation of the sample.
For $H=7 \; \rm T$, the data span the range $-0.1<t<0.1$.
For lower fields, the range is appropriately smaller.  For this
scaling plot we used the $H=0$ random-exchange exponent $\alpha ^*= -0.10$.
}
\end{figure}

\end{document}